\documentclass{moriond}

\def\Ap{\ensuremath{A^{\prime}}}
\def\nue{\ensuremath{\nu_e}}
\def\numu{\ensuremath{\nu_\mu}}

\def\nuebar{\ensuremath{\bar{\nu}_e}}
\def\numubar{\ensuremath{\bar{\nu}_\mu}}

\def\epem{\ensuremath{e^+e^-}}
\def\sqrts{\ensuremath{\sqrt{s}}}
\def\ifb{\ensuremath{\mathrm{fb}^{-1}}}

\newcommand{\Photo}{}

\begin{document}
\vspace*{4cm}
\title{Latest Results from the FASER Experiment}

\author{Shunliang~Zhang and Zhen~Hu on behalf of the FASER Collaboration}

\address{Department of Physics, Tsinghua University,
Beijing 100084, China}

\maketitle\abstracts{We present the latest physics results from the FASER
experiment at the LHC.
Using $pp$ collision data at $\sqrts = 13.6$~TeV collected during LHC Run~3,
FASER reports new results on four fronts:
a search for dark photons with an improved analysis strategy using 177~\ifb of data,
yielding world-leading exclusion limits;
neutrino cross section measurements and the first search for charm hadron
production in neutrino interactions, both using the FASER$\nu$ emulsion
detector with a 681~kg tungsten target and 9.5~\ifb of 2022 data;
the first observation of \nue in the FASER electronic
 detector at $5.5\sigma$ using 176.8~\ifb of data;
and the first double-differential measurement of \numu interactions as
a function of energy and rapidity with 186~\ifb of data.}

\section{Introduction}

The ForwArd Search ExpeRiment (FASER)~\cite{faser_det} is a detector at
the LHC located in the TI12 tunnel approximately 480~m downstream of
the ATLAS interaction point (IP1), directly on the proton--proton collision
axis line of sight (LoS). Only neutrinos and muons can traverse the $\sim$100~m
of rock between IP1 and FASER; most other Standard Model particles are
deflected by LHC magnets or absorbed by the intervening shielding.
FASER pursues two complementary physics programmes: the search for light,
long-lived beyond-the-Standard-Model (BSM) particles, and the study of
high-energy collider neutrinos in the previously unexplored TeV energy
regime~\cite{faser_nu_obs}.

Since the start of LHC Run~3 in 2022, FASER has operated with greater
than 97\% data-taking efficiency, accumulating more than 311~\ifb
of $pp$ collision data by early 2026. The results presented here use
up to 186~\ifb of electronic-detector data from 2022--2024, and
9.5~\ifb of emulsion data from 2022.

\section{The FASER Detector}

The FASER detector~\cite{faser_det} consists of several subsystems
arranged along the collision axis from IP1. Upstream, the FASER$\nu$
sub-detector is a passive stack of 730 interleaved layers of 1.1~mm-thick
tungsten plates and emulsion films, with a total mass of 1.1~tonnes. It
provides sub-micrometre position resolution for identifying neutrino
interaction vertices and classifying neutrino flavours.

The electronic detector downstream of FASER$\nu$ includes: a front
scintillator veto system, an interface tracker station (IFT), a 1.5~m
decay volume surrounded by a 0.57~T dipole magnet, a timing scintillator,
a tracking spectrometer with three stations of ATLAS SCT silicon strip
modules and two further 0.57~T dipole magnets for charge-sign and momentum
measurements, a preshower scintillator system, and an EM calorimeter made
of four LHCb ECAL modules. In early 2024, a dual-readout system for the
calorimeter was installed and a sixth veto layer was added, further
improving background rejection.

\section{Search for Dark Photons}

Dark photons ($\Ap$), arising from a kinetic mixing of a hidden $U(1)$
gauge boson with Standard Model photons, are a well-motivated BSM
benchmark. FASER is sensitive to dark photons that decay to $\epem$
pairs hundreds of metres from IP1, exploiting the nearly background-free
far-forward environment.

A new search~\cite{faser_dp26} uses the full Run~3 electronic dataset of
177~\ifb and introduces significant improvements over the previous
analysis~\cite{faser_dp_prev} (27~\ifb). Two orthogonal signal
regions are defined. The first, the ``One Track Signal Region,'' relaxes
the original two-track requirement to at least one good fiducial track,
recovering signal events where the two tracks are unresolvable due to
large boost or where EM shower activity produces additional tracks.
The second, the ``Segment Signal Region,'' uses reconstructed track
segments in individual tracking stations to gain sensitivity to $\Ap$
decays inside the spectrometer, extending the effective fiducial decay
volume by 2.5~m. Together, these two signal regions improve the
sensitivity by more than a factor of two relative to the previous
strategy.

The total expected background is $0.077^{+0.028}_{-0.042}$ events,
dominated by high-energy neutrino interactions. Zero events are observed
in both signal regions. The result sets world-leading exclusion limits
on dark photons with couplings $\epsilon \sim 10^{-5}$--$10^{-4}$ and
masses $\sim$10--150~MeV at 90\% confidence level (Fig.~\ref{fig:dp}).

\begin{figure}[h]
  \begin{minipage}{0.55\linewidth}
    \centering
    \includegraphics[height=4.8cm]{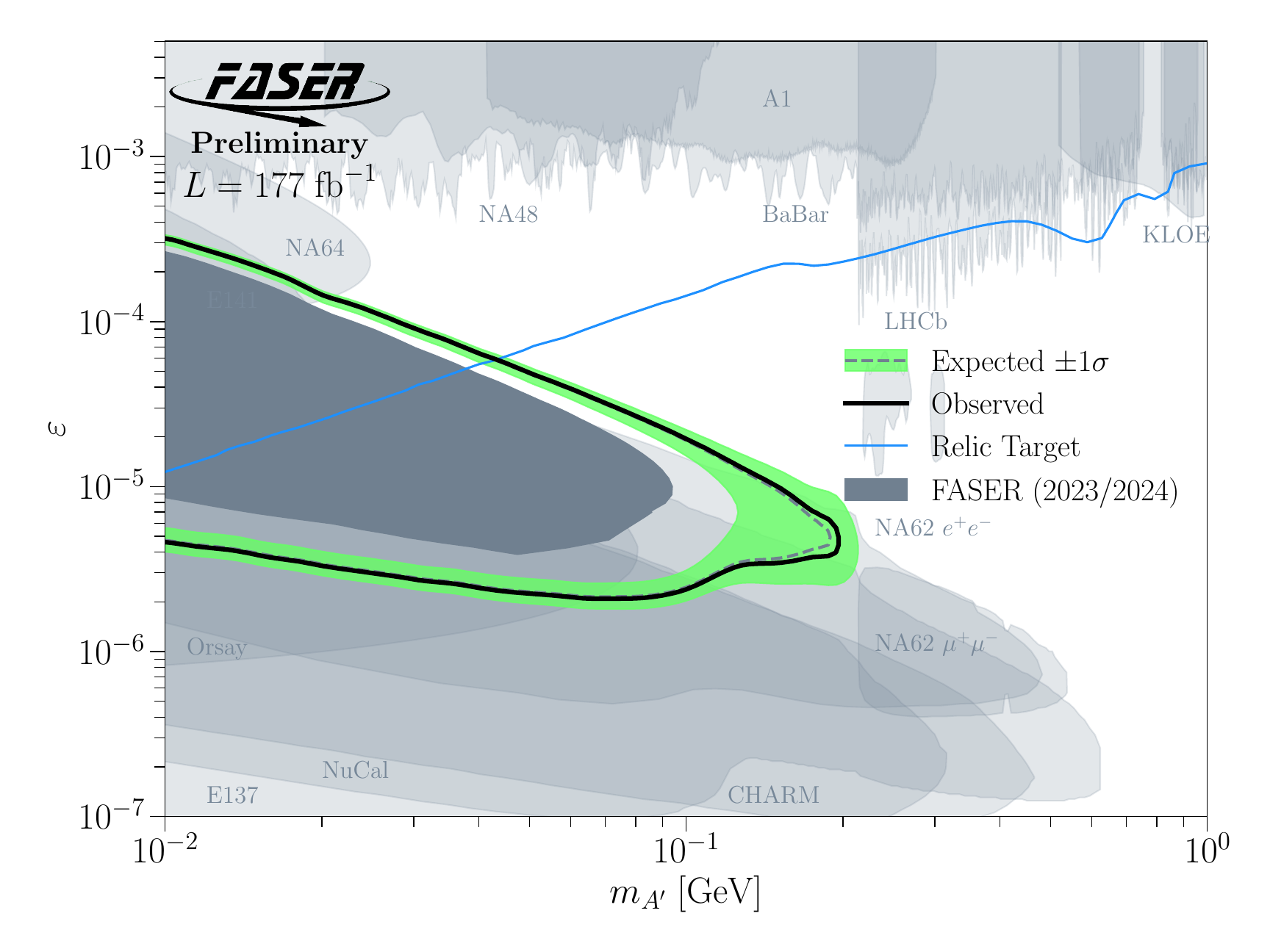}
  \end{minipage}
  \hfill
  \begin{minipage}{0.45\linewidth}
    \caption[Dark photon exclusion limits from FASER with 177 fb$^{-1}$]
    {Expected and observed 90\% CL exclusion limits on dark photon
    coupling $\epsilon$ versus mass $m_{\Ap}$ from the combined One~Track
    and Segment signal regions with 177~\ifb.~\cite{faser_dp26}
    The blue line indicates the region of parameter space that yields
    the correct dark matter relic density.
    The previous two-track analysis sensitivity is shown for comparison.}
    \label{fig:dp}
  \end{minipage}
\end{figure}

\section{Neutrino Physics}

\subsection{FASER\texorpdfstring{$\nu$}{nu} Emulsion Detector Results}

\paragraph{Updated neutrino cross section measurements.}
A new FASER$\nu$ analysis~\cite{faser_emu26} uses 9.5~\ifb of 2022
data with an expanded target mass of 681.1~kg, a factor of 2.2 increase
over the previous result in 2025. Candidate $\numu$ and $\nue$ charged-current (CC)
interactions are identified by reconstructing neutral primary vertices
with associated lepton tracks.

For the first time, the muon neutrino energy is reconstructed using a
regression boosted decision tree (BDT)~\cite{faser_mcs} that exploits the
muon momentum (measured via multiple Coulomb scattering with 20--40\%
resolution) together with the hadronic activity at the vertex. The
$\numu\,{+}\,\numubar$ cross section is measured as a function of energy
in two bins, and the $\nue\,{+}\,\nuebar$ cross section is measured in a
single bin. In total, 33 $\numu$ CC and 7 $\nue$ CC candidate events are
selected. All results are consistent with Standard Model predictions
and represent the most precise neutrino cross section measurements at
TeV energies to date (Fig.~\ref{fig:xs}, left).

\paragraph{First search for neutrino-induced charm production.}
Using the same dataset, FASER has performed the first search for charm
hadron production in high-energy neutrino CC interactions~\cite{faser_charm26}.
Dedicated tools reconstruct secondary displaced vertices from charged
charm decays ($D^0$, $D^\pm$, $D_s^\pm$, $\Lambda_c^\pm$), validated
with Monte Carlo simulation. A multivariate analysis using Lorentz-invariant
kinematic variables discriminates charm decays from hadronic interaction
backgrounds. The analysis is applied to 40 neutrino interaction candidates;
a partial unblinding of the data has been performed, with full unblinding
expected in the near future.

\begin{figure}[t]
  \begin{minipage}{0.66\linewidth}
\centerline{\includegraphics[height=5cm,width=\linewidth,keepaspectratio]{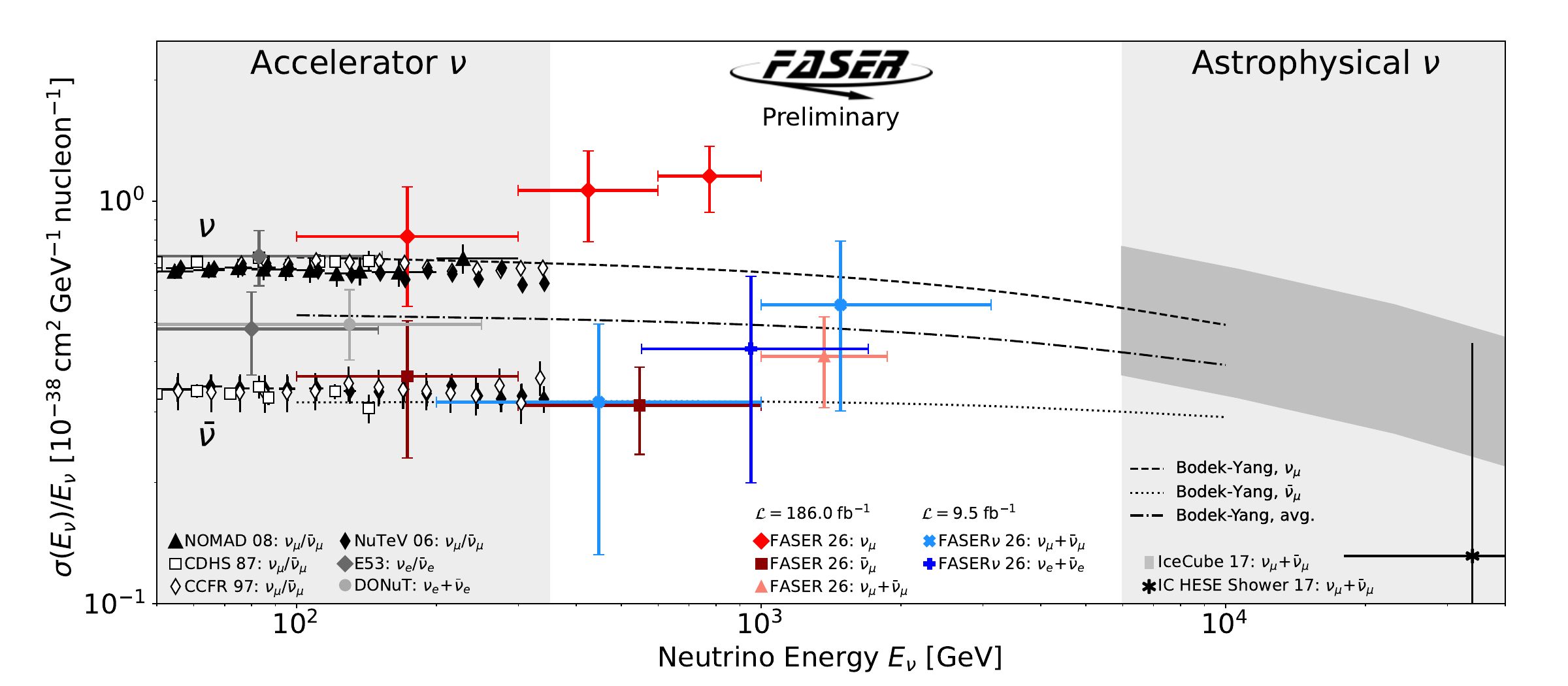}}
  \end{minipage}
  \begin{minipage}{0.34\linewidth}
\centerline{\includegraphics[height=5cm,width=\linewidth,keepaspectratio]{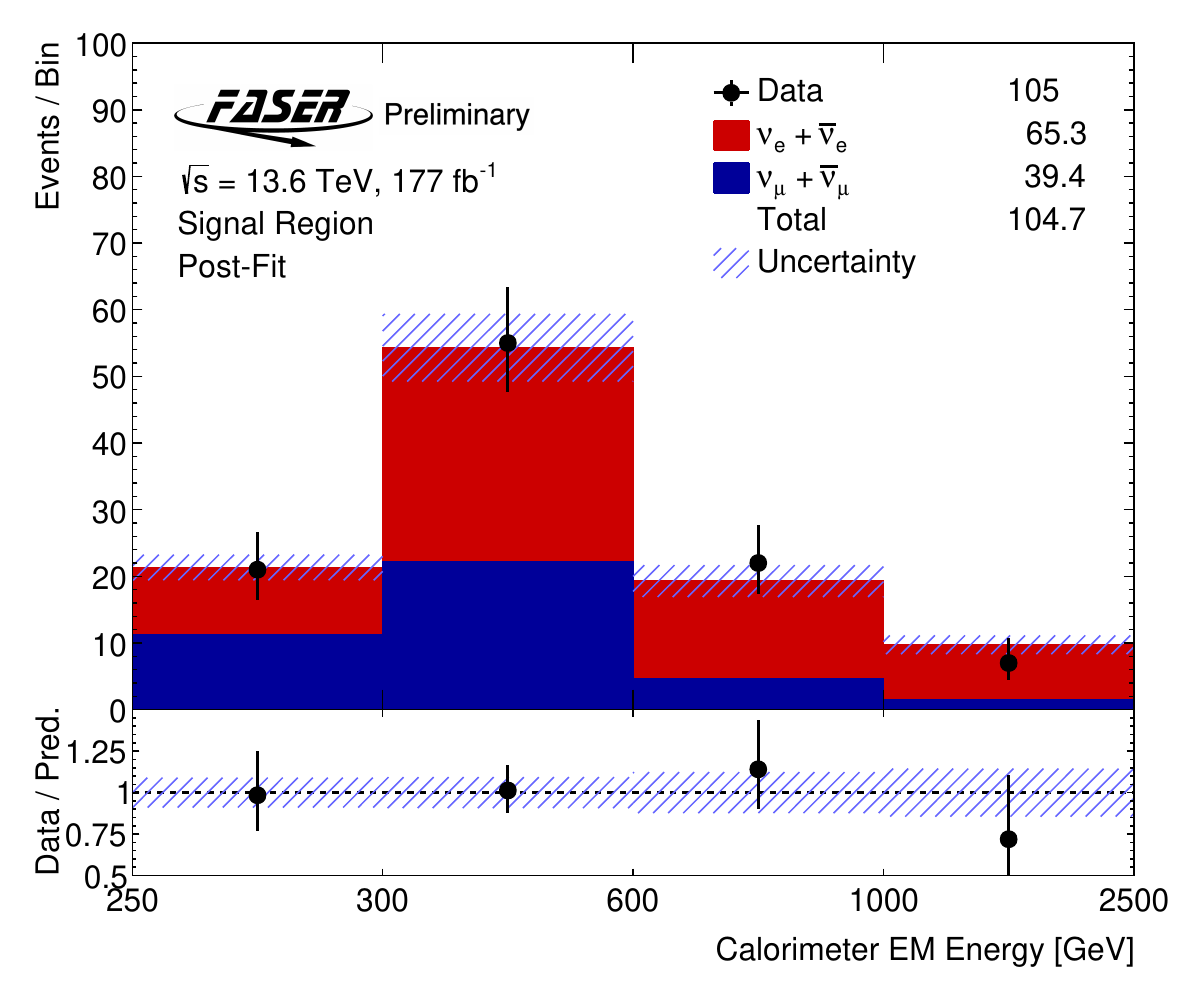}}
  \end{minipage}
  \hfill

  \caption[Neutrino cross sections and $\nue$ observation in FASER]
  {
  Left: Measured $\numu$ and $\nue$ CC neutrino--nucleon cross sections
  from the FASER$\nu$ emulsion detector~\cite{faser_emu26}
  and the FASER electronic detector~\cite{faser_numu_dd26},
  compared with theoretical predictions.
  Right: Post-fit calorimeter EM energy distribution in the $\nue$ signal
  region, showing an excess of $65\pm12$ events above
  background~\cite{faser_nue_calo26}.
  }
  \label{fig:xs}
\end{figure}

\subsection{Electron Neutrino Observation in the Electronic Detector}

FASER has observed \nue in the EM calorimeter using the
2022--2024 datasets of 176.8~\ifb.~\cite{faser_nue_calo26}
The \nue CC interactions produce large, contained EM showers with little
upstream activity. The signal region requires no signal in any veto
scintillator and a calorimeter EM energy $E_\mathrm{calo} > 250$~GeV,
divided into four energy bins.

The dominant background from $\numu$ interactions is estimated using a
data-driven approach based on FASER's measurement of the $\numu$
interaction rate as a function of energy~\cite{faser_numu_xs}, extrapolated
to the signal region via a simulation-derived migration matrix. An excess
of $65 \pm 12$ events above the background-only expectation is observed,
consistent with the expected electron-neutrino signal of $42 \pm 27$.
The background-only hypothesis is rejected at 5.5 standard deviations,
constituting the first observation of $\nue$ with the FASER electronic
detector (Fig.~\ref{fig:xs}, right).

\subsection{Double-Differential Muon Neutrino Measurement in the Electronic Spectrometer}

The first double-differential measurement of $\numu$ interactions as
simultaneous functions of neutrino energy $E$ and rapidity $y$ has been
performed~\cite{faser_numu_dd26}, using 186~\ifb of data.
CC $\numu$ interactions are selected by reconstructing muon tracks in the
FASER spectrometer originating from the FASER$\nu$ tungsten target; the
ratio $q_\mu/p_\mu$ at detector level is used as a proxy for the neutrino
rapidity. A total of $766.8 \pm 29.6$ interaction events are observed
after background subtraction.

The events are unfolded into a 100~mm-radius cylindrical fiducial volume
within the FASER$\nu$ target, yielding the interaction rate as a function
of both $E$ and $y$ simultaneously. The one-dimensional cross section is shown in Fig~\ref{fig:xs}, left. This provides new constraints
on forward QCD processes, charm production, and neutrino flux modelling
relevant to atmospheric neutrinos.

\section{Outlook}

Work is ongoing to combine the spatial resolution of the FASER$\nu$
emulsion detector with the momentum and charge-sign measurements of the
electronic spectrometer by matching tracks across the two subsystems.
This emulsion--spectrometer matching will enable charge-sign-tagged
neutrino cross section measurements, providing complementary information
to the standalone emulsion and electronic analyses.

Two new off-axis detectors were successfully installed at FASER in
January 2026 and are currently taking LHC Run~3 data.
An analog hadronic calorimeter (AHCAL), originally built as a prototype
for the CEPC hadronic calorimeter and subsequently repurposed for use
at FASER, aims to detect neutrinos originating from forward charm
production, probing gluon PDFs at $x \sim 10^{-6}$--$10^{-7}$ and
providing direct input to key physics processes at future 100~TeV
colliders. FASERCal, based on the T2K SuperFGD technology of scintillating
cubes, complements the AHCAL measurements with an independent detection
technology. Both detectors also serve as demonstrators for the HL-LHC Run~4
detector programme, for which FASER has been approved and several
on-axis and off-axis detector configurations are under
consideration~\cite{faser_run4}.

\section{Conclusions}

The FASER experiment has presented a rich set of new results at Moriond
2026. The dark photon search with an expanded analysis strategy sets the
world's strongest limits for masses of 10--150~MeV. In the neutrino
sector, the FASER$\nu$ emulsion detector delivers updated $\nue$ and
$\numu$ cross section measurements with a substantially enlarged target,
while the first search for neutrino-induced charm production at TeV
energies has been initiated. With the Run~3 2022--2024 electronic dataset, the
first observation of $\nue$ interactions in the FASER electronic detector
is achieved at $5.5\sigma$ significance, and the first double-differential
measurement of $\numu$ interactions rounds out a comprehensive suite of
new neutrino results. These achievements demonstrate FASER's unique
capabilities in the far-forward regime and pave the way for further
measurements as Run~3 continues.

\section*{Acknowledgments}
We thank CERN for the excellent performance of the LHC and the technical and administrative
staff members at all FASER institutions. We acknowledge support by the NSFC grant No. W2511007 and the Tsinghua University Initiative Scientific Research Program.

\section*{References}
\bibliography{moriond}

@article{faser_det,
  author        = "{FASER Collaboration}",
  title         = "{The FASER Detector}",
  journal       = "JINST",
  volume        = "19",
  pages         = "P05066",
  year          = "2024",
  doi           = "10.1088/1748-0221/19/05/P05066"
}

@article{faser_nu_obs,
  author        = "{FASER Collaboration}",
  title         = "{First Direct Observation of Collider Neutrinos with FASER at the LHC}",
  journal       = "Phys. Rev. Lett.",
  volume        = "131",
  pages         = "031801",
  year          = "2023",
  doi           = "10.1103/PhysRevLett.131.031801"
}

@article{faser_numu_xs,
  author        = "{FASER Collaboration}",
  title         = "{First Measurement of the Muon Neutrino Interaction Cross Section and Flux as a Function of Energy at the LHC with FASER}",
  journal       = "Phys. Rev. Lett.",
  volume        = "134",
  pages         = "211801",
  year          = "2025",
  doi           = "10.1103/PhysRevLett.134.211801"
}

@article{faser_dp_prev,
  author        = "{FASER Collaboration}",
  title         = "{Search for dark photons with the FASER detector at the LHC}",
  journal       = "Phys. Lett. B",
  volume        = "848",
  pages         = "138378",
  year          = "2024",
  doi           = "10.1016/j.physletb.2023.138378",
  eprint        = "2308.05587",
  archivePrefix = "arXiv",
  primaryClass  = "hep-ex"
}

@article{faser_dp26,
  author        = "{FASER Collaboration}",
  title         = "{New Results from a Search for Dark Photons with the FASER Detector at the LHC}",
  note          = "CERN-FASER-CONF-2026-001",
  year          = "2026"
}

@article{faser_emu26,
  author        = "{FASER Collaboration}",
  title         = "{Cross section measurements of high-energy electron and muon neutrino interactions with FASER's emulsion detector at the LHC}",
  note          = "CERN-FASER-CONF-2026-002",
  year          = "2026"
}

@article{faser_charm26,
  author        = "{FASER Collaboration}",
  title         = "{First search for neutrino-induced charm hadrons with FASER's emulsion detector at the LHC}",
  note          = "CERN-FASER-CONF-2026-003",
  year          = "2026"
}

@article{faser_nue_calo26,
  author        = "{FASER Collaboration}",
  title         = "{Measurement of High-Energy Electron Neutrino Interactions with the FASER Calorimeter at the LHC}",
  note          = "CERN-FASER-CONF-2026-004",
  year          = "2026"
}

@article{faser_numu_dd26,
  author        = "{FASER Collaboration}",
  title         = "{First Measurement of Muon Neutrinos as a Function of Energy and Rapidity with FASER}",
  note          = "CERN-FASER-CONF-2026-005",
  year          = "2026"
}

@article{faser_mcs,
  author        = "{FASER Collaboration}",
  title         = "{Momentum Measurement of Charged Particles in FASER's Emulsion Detector at the LHC}",
  note          = "arXiv:2602.17575",
  year          = "2026",
  eprint        = "2602.17575",
  archivePrefix = "arXiv",
  primaryClass  = "physics.ins-det"
}

@article{faser_run4,
  author        = "{FASER Collaboration}",
  title         = "{Prospects and Opportunities with an upgraded FASER Neutrino Detector during the HL-LHC era: Input to the EPPSU}",
  note          = "arXiv:2503.19775",
  year          = "2025",
  eprint        = "2503.19775",
  archivePrefix = "arXiv",
  primaryClass  = "hep-ex"
}

\end{document}

